\documentclass[letterpaper, 10 pt, conference]{ieeeconf}  

\IEEEoverridecommandlockouts                              
\usepackage{graphicx} 
\usepackage{amsmath} 
\usepackage{amssymb}  
\usepackage{booktabs}
\usepackage{subfigure}
\usepackage{siunitx}

\title{\LARGE \bf
Robust Air Data Sensor Fault Diagnosis With Enhanced Fault Sensitivity using Moving Horizon Estimation  
}

\author{Yiming Wan, Tamas Keviczky, and Michel Verhaegen 
\thanks{This work has received funding from the European Union's Seventh Framework Programme (FP7-RECONFIGURE/2007-2013) under grant agreement no. 314544.}
\thanks{Yiming Wan, Tamas Keviczky, and Michel Verhaegen are with Delft Center for Systems and Control, Delft University of Technology, 2628CD, The Netherlands.
        {\tt\small y.wan@tudelft.nl, T.Keviczky@tudelft.nl, m.verhaegen@tudelft.nl}}%
}

\DeclareMathOperator*{\argmin}{\arg\,\min}

\newtheorem{thm}{Theorem}

\begin{document}

\maketitle
\thispagestyle{empty}
\pagestyle{empty}

\begin{abstract}
This paper investigates robust fault diagnosis of multiple air data sensor faults in the presence of winds. The trade-off between robustness to winds and sensitivity to faults is challenging due to simultaneous influence of winds and latent faults on monitored sensors. 
Different from conventional residual generators that do not consider any constraints, we propose a constrained residual generator using moving horizon estimation. 
The main contribution is improved fault sensitivity by exploiting known bounds on winds in residual generation. By analyzing the Karush-Kuhn-Tucker conditions of the formulated moving horizon estimation problem, it is shown that this improvement is attributed to active inequality constraints caused by faults. 
When the weighting matrices in the moving horizon estimation problem are tuned to increase robustness to winds, its
fault sensitivity does not simply decrease as one would expect in conventional unconstrained residual generators. Instead, its fault sensitivity increases
when the fault is large enough to activate some inequality constraints. 
This fault sensitivity improvement is not restricted to this particular application, but can be achieved by any general moving horizon estimation based residual generator.
A high-fidelity Airbus simulator is used to illustrate the advantage of our proposed approach in terms of fault sensitivity.
\end{abstract}

\section{INTRODUCTION}
During aircraft operations, air data sensor measurements are fed into the flight control computer to calculate the flight control law, thus it is critical to identify any air data sensor faults \cite{Goup2014,Goup2015}. 
The industrial state-of-the-art for sensor fault detection and isolation (FDI) relies on triplex hardware redundancy, and performs a majority voting scheme to select the accurate measurements and discard any failed sources \cite{Goupil2011}. This scheme works well if only one sensor source becomes faulty, but it would be inadequate to address simultaneous multiple sensor faults within the triplex redundancy.
As currently investigated in the RECONFIGURE project \cite{Goup2014,Goup2015}, one possibility to extend guidance and control functionalities for future aircraft could be the incorporation of analytical redundancy to detect, isolate and estimate sensor faults without adding new sensors. 

In the analytical redundancy based FDI technique, an important method is the residual-based approach using various types of linear and nonlinear observers or Bayesian filters \cite{Ding2013,Marz2012}. 
A crucial issue with any FDI scheme in aircraft applications is how to simultaneously maintain its robustness to wind disturbances and optimize its fault sensitivity. 
In \cite{Patton1992}, a robust fault detection approach based on eigenstructure assignment was proposed for faulty sensors of jet engines.
In \cite{Eykeren2014,Hansen2014}, an extended Kalman filter (EKF) based FDI method was proposed based on the assumption of constant winds. This simplified assumption limits their applicability to situations without wind shear.
Without any limiting assumption about wind dynamics, the disturbance decoupling method based on differential geometry was used in \cite{Cast2010,Cast2014} to perfectly decouple wind effect in the generated residual signal. 

This paper focuses on fault diagnosis of air data sensors in the presence of winds. 
It is challenging because wind disturbances and latent sensor faults simultaneously affect some air data measurements \cite{Goup2014,Goup2015}. In this case, the disturbance decoupling method adopted in \cite{Cast2010,Cast2014} cannot be applied, since decoupling the wind effect in the residual signal would also decouple the corresponding air data sensor faults to be detected. As an alternative to perfect disturbance decoupling, we may increase robustness to disturbances at the cost of reducing sensitivity to faults \cite{Ding2013}.

Motivated by the above challenge, we propose a constrained residual generator based on a system model augmented with wind dynamics. 
The dynamic relations between wind speed and acceleration are described by a first-order integrating model, 
and the information about bounds on wind speed and acceleration \cite{Goup2014,Goup2015} is exploited in a constrained moving horizon estimation (MHE) problem formulation which is known for its capacity to handle constraints \cite{Rawl2006}. 
By analyzing the Karush-Kuhn-Tucker (KKT) conditions of the formulated MHE problem, it is shown that
the main advantage of incorporating constraints in residual generation is the improvement of fault sensitivity attributed to the active constraints caused by faults. 
When tuning the weighting matrices in the MHE problem to increase disturbance robustness, 
its fault sensitivity would increase, rather than decrease as one would expect in conventional unconstrained residual generators, if the fault is large enough to activate some constraints. It is worth noting that the fault sensitivity improvement is not restricted to this particular application, but can be achieved by any general MHE based residual generator.


This paper is organized as follows. Section \ref{sect:prob} describes the system model.
Then the proposed FDI scheme is given in Section \ref{sect:FTE}. Section \ref{sect:fsen_MHEGR} explains how active constraints caused by faults contribute to improving fault sensitivity. Simulation examples on the Airbus simulator are detailed in Section \ref{sect:sim}. Conclusions and future work are discussed in Section \ref{sect:con}.

\section{SYSTEM DESCRIPTION AND PROBLEM FORMULATION}
\label{sect:prob}
In order to enhance triplex monitoring which can address only one single fault in the redundant sensors, we exploit analytical redundancy to detect and isolate simultaneous multiple angle-of-attack (AOA) and calibrated airspeed (VCAS) sensor faults in this paper.


Since only longitudinal dynamics is investigated in this paper, the following model is adopted:
\begin{equation}\label{eq:ct_ss}
\left\{
\begin{aligned}
{\mathbf{\mathbf{\dot \alpha}}} (t) &= {f} \left( \alpha (t), \varTheta (t) \right) + {u_{\alpha}}(t)  \\
{\mathbf{y}} (t) &= h \left( \alpha (t), \mathbf{w} (t), \varTheta (t) \right) \\
{\mathbf{y}}_m (t) &= {\mathbf{y}} (t) + \mathbf{n}(t)
\end{aligned}
\right.
\end{equation}
with the definitions 
$\varTheta = \left[ \begin{matrix}
V_g & \theta & q & n_x & n_z & z
\end{matrix} \right]^\mathrm{T}$,  
$\mathbf{w} = \left[ \begin{matrix}
W_x & W_z
\end{matrix} \right]^\mathrm{T}$,  
$\mathbf{y} = \left[ \begin{matrix}
\alpha & V_z & V_{c}
\end{matrix} \right]^\mathrm{T}$,
$\mathbf{n} = \left[ \begin{matrix}
n_{\alpha} & n_{vz} & n_{vc}
\end{matrix} \right]^\text{T}$.
The system outputs ${\mathbf{y}} (t)$ include AOA $\alpha$, vertical speed $V_z$, and VCAS $V_c$. 
The output measurements $\mathbf{y}_m (t)$ are corrupted with measurement noises $\mathbf{n}(t)$.
$W_x$ and $W_z$ represent horizontal and vertical wind speeds, respectively. 
The model parameter $\varTheta$ consists of ground speed $V_g$, pitch angle $\theta$, pitch rate $q$, horizontal load factor $n_x$, vertical load factor $n_z$, and altitude $z$, which are all measurable.
$u_\alpha$ is the input noise accounting for the model mismatch.
The output equations in (\ref{eq:ct_ss}) for $V_z$ and 
$V_c$ are
\begin{equation}\label{eq:hvc}
V_z = h_{vz} (\alpha, \mathbf{w}, \varTheta)  \text{ and }
V_{c} = h_{vc} (\alpha, \mathbf{w}, \varTheta), 
\end{equation}
respectively. For each redundant AOA sensor measurement $\alpha_{m}^{(i)}$ or VCAS sensor measurement $V_{m,c}^{(i)}$, $i = 1,2,3$, the latent sensor faults $f_{\alpha}^{(i)}$ and $f_{vc}^{(i)}$ are additive, i.e.,
\begin{equation}\label{eq:hvc_f}
\alpha_{m}^{(i)} = \alpha + f_{\alpha}^{(i)} + n_{\alpha}^{(i)},\;
V_{c,m}^{(i)} = V_c + f_{vc}^{(i)} + n_{vc}^{(i)}.
\end{equation}

The system model (\ref{eq:ct_ss}) is adopted due to several considerations: a) it avoids other air data measurements which are considered as unreliable in the presence of AOA/VCAS sensor faults \cite{Goup2015}, and involves only inertial sensors; b) it includes no aerodynamic parameters, avoiding the issue of robustness to uncertain aerodynamic parameters; c) its low state dimensions are attractive for real-time computation. More details about the model (\ref{eq:ct_ss}) are given in Appendix \ref{app:model}.

Besides the approximately zero-mean measurement noises $\mathbf{n}(t)$ and input noise $u_\alpha (t)$, the main source of uncertainty is the wind disturbance $\mathbf{w}(t)$ which should be distinguished from sensor faults in the considered FDI problem.
Note that there is no direct wind effect on the AOA measurements in the model (\ref{eq:ct_ss}), thus we can easily generate a residual for AOA FDI by decoupling the wind effect from verticle speed $V_z$ and VCAS. In contrast,
the task of VCAS FDI is more challenging, because VCAS measurements are simultaneously affected by winds and latent sensor faults, as shown in (\ref{eq:hvc}) and (\ref{eq:hvc_f}). 
In this case, perfect disturbance decoupling would lead to complete loss of sensitivity to fault. 
A remedy could be to introduce a larger detection threshold at the cost of reduced fault sensitivity. In order to address this above challenge, the basic idea of our proposed solution is to augment the model (\ref{eq:ct_ss}) with wind dynamics and to adopt constrained residual generation by exploiting known bounds on wind speed and acceleration.

\section{FAULT-TOLERANT MOVING HORIZON ESTIMATION SCHEME}\label{sect:FTE}
\subsection{FDI Scheme}\label{sect:FTEscheme}
\begin{figure}[!h]
	\centering
	\includegraphics[width=0.7\linewidth]{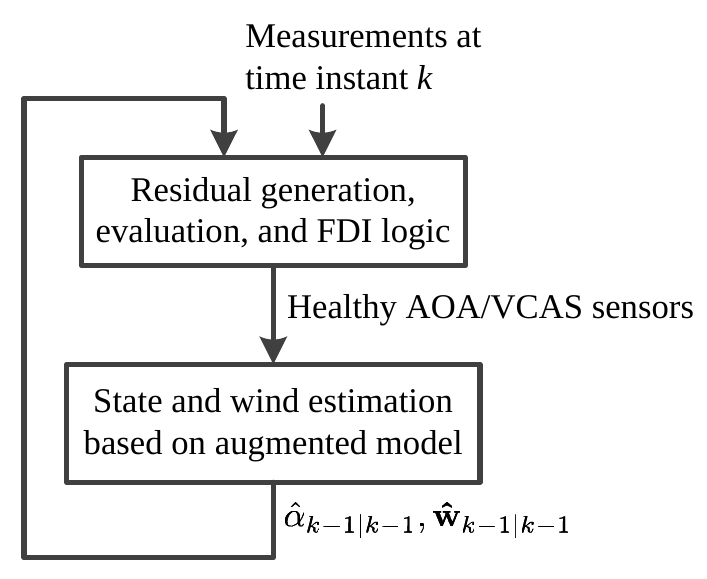}
	\caption{Fault detection and isolation scheme}
	\label{fig:FTE_scheme}
\end{figure}

As depicted in Figure \ref{fig:FTE_scheme}, our FDI scheme consists of two consecutive steps: a) isolating faulty AOA/VCAS sensors based on generated residual signals; b) estimating states and winds after removing faulty sensors.
The residual signals for FDI are generated as the difference between the AOA/VCAS measurements and their one-step-ahead predictions, i.e.,
\begin{equation}\label{eq:respred_vc}
\begin{aligned}
r_{\alpha,k}^{(i)} &= \alpha_{m,k}^{(i)} - \hat \alpha_{k|k-1}, \\
r_{vc, k}^{(i)} &= V_{c,m,k}^{(i)} - \hat V_{c,k|k-1}, \; i=1,2,3.
\end{aligned}
\end{equation}
Here, the index $k$ denotes the samples at time instant $t_k$. The one-step ahead predictions $\hat \alpha_{k|k-1}$ and $\hat V_{c,k|k-1}$ are computed from the filtered estimates $\hat \alpha_{k-1|k-1}$ and $\mathbf{\hat w}_{k-1|k-1}$ based on the model (\ref{eq:ct_ss}) and the assumed wind dynamics given latter.
The residual signals are evaluated by their root mean square (RMS) values over a sliding window \cite{Ding2013}:
\begin{equation*}
J_{\star,k}^{(i)} = \sqrt{ \frac{1}{N_{\text{eval}}} \sum_{j=k-N_{\text{eval}}+1}^{k} (r_{\star,j}^{(i)})^2 }
\end{equation*}
where $\star$ represents ``$\alpha$'' or ``$vc$'', $N_{\text{eval}}$ is the length of residual evaluation window. 
The adopted fault detection logic is 
\begin{equation}\label{eq:FDlogic}
\left\{ \begin{array}{ll}
J_{\star,k}^{(i)} > J_{\star, \mathrm{th}} & \Rightarrow i^{\text{th}} \text{ AOA or VCAS sensor is faulty} \\
J_{\star,k}^{(i)} \le J_{\star, \mathrm{th}} & \Rightarrow i^{\text{th}} \text{ AOA or VCAS sensor is fault-free}.
\end{array}
\right.
\end{equation}
The threshold can be obtained by hypothesis test based on the approximated distribution of the fault-free residual signal, or empirical Monte Carlo simulations based on normal historical measurement data, to avoid false alarms \cite{Ding2013, Hansen2014}.

State and wind estimation in Figure \ref{fig:FTE_scheme} incorporates the following first-order integrating model to represent wind dynamics:
\begin{equation}\label{eq:wind_dyn}
\mathbf{\dot x}_d = \mathbf{u}_d \text{ with }
\mathbf{x}_d = \left[ \begin{matrix}
W_x & W_z 
\end{matrix}  \right]^{\mathrm{T}}, \;
\mathbf{u}_d = \left[ \begin{matrix}
u_{d,x} & u_{d,z}
\end{matrix} \right]^{\mathrm{T}},
\end{equation}
%
$u_{d,x}$ and $u_{d,z}$ represent horizontal and vertical wind acceleration. 
This assumed wind dynamics was proposed for aircraft control in wind shear \cite{Mulg1996}.
It is exploited in our paper, however, for the purpose of FDI and estimation.

\subsection{Overview of MHE}\label{sect:overviewMHE}
The MHE approach is adopted for state and wind estimation in our FDI scheme. 
Besides addressing nonlinearity and robustness to initial errors \cite{Rao2003}, the employment of MHE enables enhancement of 
fault sensitivity by exploiting known bounds on wind speeds and accelerations, which will be explained in Section \ref{sect:fsen_MHEGR}.
A brief overview of the MHE technique is first given below.

The MHE framework builds on the discrete-time approximation of the continuous-time model (\ref{eq:ct_ss}) and the assumed wind dynamics (\ref{eq:wind_dyn}):
\begin{subequations}\label{eq:dt_ss}
	\begin{align}
	\alpha_{k+1} &= \alpha_k + t_s f(\alpha_k, \varTheta_k) + t_s {u}_{\alpha,k},    \label{eq:dt_longdyn} \\
	\mathbf{x}_{d,k+1} &= \mathbf{x}_{d,k} + t_s \mathbf{u}_{d,k}, \label{eq:dt_winddyn} \\
	\mathbf{\bar y}_{m,k} &= h(\alpha_k, \mathbf{x}_{d,k}, \varTheta_k) + \mathbf{\bar n}_{k},   \label{eq:dt_longoutput}
	\end{align}
\end{subequations}
where $t_s$ represents the sampling interval, (\ref{eq:dt_longdyn}) 
and (\ref{eq:dt_winddyn}) are obtained via approximated numerical integration applied to (\ref{eq:ct_ss}) and (\ref{eq:wind_dyn}). In (\ref{eq:dt_longoutput}), all redundant AOA and VCAS measurements identified as healthy by the FDI step (see Figure \ref{fig:FTE_scheme}) are merged into two single components in the output vector $\mathbf{\bar y}_{m,k}$, respectively, i.e.,
\begin{equation}\label{eq:ymeas}
\mathbf{\bar y}_{m,k} = \left[ \begin{matrix}
\frac{\sum_{j \in \mathcal{H}_\alpha} \alpha_{m,k}^{(j)}}{N_\alpha}  & V_{z,m,k} 
& \frac{\sum_{j \in \mathcal{H}_{vc}} V_{c,m,k}^{(j)}}{N_{vc}} 
\end{matrix} \right]^\text{T},
\end{equation}
where $\mathcal{H}_\alpha$ and $\mathcal{H}_{vc}$ represent the sets of AOA and VCAS sensors identified as healthy, $1 \leq N_{\alpha} \leq 3$ and $1 \leq N_{vc} \leq 3$ are the number of sensors in the sets $\mathcal{H}_\alpha$ and $\mathcal{H}_{vc}$, respectively. $\mathbf{\bar n}_k$ is defined similarly to (\ref{eq:ymeas}).

Given a moving horizon including $N$ samples of system output measurements $\{ \mathbf{\bar y}_{m,l}, \mathbf{\bar y}_{m,l+1}, \dots, \mathbf{\bar y}_{m,k} \}$ ($l = k-N+1$) at time instant $k$, the MHE problem is formulated as 
\begin{subequations}\label{eq:MHE}
\begin{align}
\mathop {\min }\limits_{\begin{smallmatrix}
							\mathbf{x}_{i}, \mathbf{u}_{i} 
						\end{smallmatrix}} \;
	& \frac{1}{2}\left\| {{\mathbf{x}_l} - {\mathbf{x}_l^-} } \right\|_{\mathbf{P}^{ - 1}}^2
	  + \frac{1}{2}\sum\limits_{i = l}^{k - 1} {\left\| \mathbf{u}_{i} \right\|_{\mathbf{Q}^{-1}}^2}   \label{eq:MHEobj} \\
	& + \frac{1}{2}\sum\limits_{i = l}^k {\left\| {\mathbf{\bar y}_{m,i} - h ({\mathbf{x}_i}, \varTheta_i) } \right\|_{\mathbf{R}^{ - 1}}^2}  \nonumber \\
\text{s.t.} \; & \mathbf{x}_{i+1} = F(\mathbf{x}_i, \mathbf{u}_i, \varTheta_i),  \label{eq:MHEcon} \\
               & \mathbf{u}^{\text{LB}} \leq \mathbf{u}_{i} \leq \mathbf{u}^{\text{UB}}, \; i=l,\dots,k-1, \nonumber \\
               & \mathbf{x}^{\text{LB}} \leq \mathbf{x}_{i} \leq \mathbf{x}^{\text{UB}}, \; i = l,\dots,k, \nonumber
\end{align}
\end{subequations}
where
\begin{equation}\label{eq:MHEdef}
\begin{aligned}
& \mathbf{x} = \left[ \begin{matrix}
\alpha \\
\mathbf{x}_d
\end{matrix} \right], 
\mathbf{u} = \left[ \begin{matrix}
u_\alpha \\
\mathbf{u}_d
\end{matrix} \right], 
\mathbf{P} = \left[ \begin{matrix}
p_\alpha & \mathbf{0}\\
\mathbf{0} & p_d \mathbf{I}_2
\end{matrix} \right], \\
&\mathbf{Q} = \left[ \begin{matrix}
q_\alpha & \mathbf{0}\\
\mathbf{0} & q_d \mathbf{I}_2
\end{matrix} \right], 
\mathbf{R} = \text{diag}(\frac{R_\alpha}{N_\alpha}, R_{vz}, \frac{R_{vc}}{N_{vc}}),
\end{aligned} 
\end{equation}
the weighting matrices $\mathbf{P}$, $\mathbf{Q}$, and $\mathbf{R}$ are all diagonal positive definite,
and the function $F(\cdot)$ in (\ref{eq:MHEcon}) represents the right-hand side of (\ref{eq:dt_longdyn}) and (\ref{eq:dt_winddyn}).
The first term of the objective function (\ref{eq:MHEobj}) is the so-called arrival cost to account for data before the current estimation horizon, where $\mathbf{x}_l^-$ represents the a priori state estimate \cite{Rawl2006}. 
In the Bayesian framework, the weighting matrices $\mathbf{P}$, $\mathbf{Q}$, and $\mathbf{R}$ in  (\ref{eq:MHEobj}) can be explained as covariance matrices, 
and the problem (\ref{eq:MHE}) formulates the maximum likelihood estimation \cite{Rawl2006}. When reliable covariance information of measurement noises and the a priori state estimates is unavailable, as in this air data sensor fault diagnosis problem, the weighting matrices can be regarded as tuning parameters. 
Note that $R_\alpha$ and $R_{vc}$ are weighting parameters for each AOA and VCAS sensors, and the weighting matrix $\mathbf{R}$ takes its form in (\ref{eq:MHEdef}) according to the definition of $\mathbf{\bar y}_{m,k}$ in (\ref{eq:ymeas}).
We will discuss how these tuning parameters affect both robustness to winds and sensitivity to faults later in Section \ref{sect:tune}.


Throughout this paper, the MHE problem (\ref{eq:MHE}) with or without inequality constraints is referred to as  constrained or unconstrained MHE (CMHE or UMHE), respectively. The benefit of incorporating constraints in residual generation will be analyzed by comparing CMHE with UMHE in terms of fault sensitivity.
Rigorous comparisons with other forms of unconstrained residual generators are out of the scope of this paper. The algorithm implemented to solve the MHE problem (\ref{eq:MHE}) adopts a real-time iteration scheme with the interior-point sequential quadratic programming strategy, with its details explained in \cite{Wan2015,Wan-IFAC2016}.

\section{FAULT SENSITIVITY OF MHE-BASED RESIDUAL GENERATOR}\label{sect:fsen_MHEGR}
In this section, we will analyze the improved fault sensitivity of CMHE based residual generator (CMHE-RG) by comparing with UMHE based residual generator (UMHE-RG), and then explain the trade-off between 
fault sensitivity and disturbance robustness when tuning the weighting matrices in the objective function (\ref{eq:MHEobj}).

Before a rigorous analysis, intuitive explanations can be given below. Sensor faults contaminate the measurements before being detected. In UMHE-RG, the state estimates are adjusted to compensate for the fault effect, thus the output residuals (\ref{eq:respred_vc}) might be still small even in the presence of faults. On the contrary, CMHE-RG respects the inequality constraints in (\ref{eq:MHEcon}) when adjusting its state estimates. When the presence of faults causes some inequality constraints to become active, the state estimates would be
restricted by the active constraints and reluctant to compensate for the fault effect, thus the generated residual signal becomes larger, i.e., more sensitive to faults. 

\subsection{Fault Sensitivity of Unconstrained-MHE-based Residual}
\label{sect:sens_UMHE}
By defining
\begin{align}
\mathbf{z}_k &= \left[ \begin{matrix}
			              \mathbf{x}_l^\mathrm{T} & \mathbf{u}_l^\mathrm{T} & \cdots & \mathbf{x}_{k-1}^\mathrm{T} & \mathbf{u}_{k-1}^\mathrm{T} &  \mathbf{x}_k^\mathrm{T}
			    \end{matrix} \right]^\mathrm{T}, \label{eq:zk} \\
\mathcal{I}_k &= \left[ \begin{matrix}
						(\mathbf{x}_l^-)^\mathrm{T} & \mathbf{0}^\mathrm{T} & \mathbf{\bar y}_{m,l}^\mathrm{T}	
						& \cdots & \mathbf{0}^\mathrm{T} & \mathbf{\bar y}_{m,k-1}^\mathrm{T} & \mathbf{\bar y}_{m,k}^\mathrm{T} \end{matrix} \right], \label{eq:Ik} \\
\mathbf{V} &= \text{diag}\left( \mathbf{P}, \mathbf{Q}, \mathbf{R}, \cdots, \mathbf{Q}, \mathbf{R}, \mathbf{R} \right), \nonumber
\end{align}
the MHE problem (\ref{eq:MHE}) can be written in the following compact form
\begin{equation}
\begin{aligned}\label{eq:MHEcompact}
\mathbf{\hat z}_k (\mathcal{I}_k) = \argmin \limits_{\mathbf{z}_k} &\; \frac{1}{2} \left\| \mathcal{I}_k - F_1 (\mathbf{z}_k) \right\|_{\mathbf{V}^{-1}}^2 \\
\text{s.t.} &\; F_2 (\mathbf{z}_k) = \mathbf{0}.
\end{aligned}
\end{equation}
The inequality constraints in (\ref{eq:MHEcon}) are omitted in this subsection, and will be discussed in Section \ref{sect:fsen_CMHE}. It can be seen that (\ref{eq:MHEcompact}) defines a function which produces the estimate $\mathbf{\hat z}_k$ from the information vector $\mathcal{I}_k$.
Since $\mathcal{I}_k$ consists of the fault-free part $\mathcal{I}_k^0$ and the sensor fault perturbation $\epsilon_k$, i.e., 
\begin{equation}\label{eq:info_vector}
\mathcal{I}_k = \mathcal{I}_k^0 + \epsilon_k
\end{equation}
with  
\begin{equation}
\epsilon_k = \left[ \begin{matrix}
\mathbf{0}^\text{T} & \mathbf{0}^\text{T} & \mathbf{f}_l^\text{T} & \cdots &
\mathbf{0}^\text{T} & \mathbf{f}_{k-1}^\text{T} & \mathbf{f}_{k}^\text{T}
\end{matrix} \right]^\text{T},
\end{equation}
fault sensitivity of the predicted signal can be analyzed via sensitivity of $\mathbf{\hat z}_k (\mathcal{I}_k^0 + \epsilon_k)$ in (\ref{eq:MHEcompact}) to the fault perturbation $\epsilon_k$.


The KKT conditions for the problem (\ref{eq:MHEcompact})
are given by 
\begin{equation}\label{eq:KKT}
\left\{
\begin{array}{l}
- \mathbf{J}_1^\mathrm{T} (\mathbf{z}_k) \mathbf{V}^{-1} \left[\mathcal{I}_k^0 + \epsilon_k - F_1 (\mathbf{z}_k) \right] + \mathbf{J}_2^\mathrm{T} (\mathbf{z}_k) \lambda = \mathbf{0}  \\
F_2 (\mathbf{z}_k) = \mathbf{0}
\end{array}
\right.
\end{equation}
where we define $\mathbf{J}_1^\mathrm{T} (\mathbf{z}_k) = \frac{\partial F_1 (\mathbf{z}_k)}{\partial \mathbf{z}_k}$ and $\mathbf{J}_2^\mathrm{T} (\mathbf{z}_k) = \frac{\partial F_2 (\mathbf{z}_k)}{\partial \mathbf{z}_k}$.
Note that $\mathbf{\hat z}_k (\mathcal{I}_k^0)$ and $\mathbf{\hat z}_k (\mathcal{I}_k^0 + \epsilon_k)$ 
are the solutions to the MHE problem (\ref{eq:MHEcompact}) in the fault-free case ($\epsilon_k = \mathbf{0}$) and the faulty case ($\epsilon_k \neq \mathbf{0}$), respectively. 
Then 
\begin{equation}\label{eq:Dz}
	\Delta \mathbf{\hat z}_k (\mathcal{I}_k^0, \epsilon_k) = \mathbf{\hat z}_k (\mathcal{I}_k^0 + \epsilon_k) - \mathbf{\hat z}_k (\mathcal{I}_k^0)
\end{equation}
should satisfy the linearized KKT conditions given below:
\begin{equation}\label{eq:KKTlin}
\left[ \begin{matrix}
\mathbf{H} & \mathbf{J}_2^\mathrm{T}  \\
\mathbf{J}_2  & \mathbf{0}
\end{matrix} \right] 
\left[ \begin{matrix}
\Delta \mathbf{\hat z}_k \\
\Delta \lambda
\end{matrix} \right] =
\left[ \begin{matrix}
\mathbf{J}_1^\mathrm{T} \mathbf{V}^{-1} \epsilon_k \\
\mathbf{0}
\end{matrix} \right].
\end{equation}
Note that the Jacobian matrices $\mathbf{J}_1$ and $\mathbf{J}_2$ in (\ref{eq:KKTlin}) are defined at $\mathbf{\hat z}_k (\mathcal{I}_k^0)$, and the Hessian matrix $\mathbf{H} = \mathbf{J}_1^\text{T} \mathbf{V} \mathbf{J}_1$ being positive definite for the considered MHE problem (\ref{eq:MHE}).
The dependence of $\Delta \mathbf{\hat z}_k$ and $\Delta \lambda$ on $\mathcal{I}_k^0$ and $\epsilon_k$ is omitted hereafter for the sake of brevity.
We assume that the linear independence constraint qualification (LICQ) and sufficient second order condition hold \cite{Noce2006}. Then the above linearized KKT system (\ref{eq:KKTlin}) can be solved by using inversion of block matrices, and we obtain
\begin{equation}\label{eq:Zsolution}
\Delta \mathbf{\hat z}_k = \mathbf{X} \mathbf{J}_1^\mathrm{T} \mathbf{V}^{-1} \epsilon_k
\end{equation}
with
\begin{equation}\label{eq:X}
\mathbf{X} = \mathbf{H}^{-1} - \mathbf{H}^{-1} \mathbf{J}_2^\mathrm{T} ( \mathbf{J}_2 \mathbf{H}^{-1} \mathbf{J}_2^\mathrm{T} )^{-1} \mathbf{J}_2 \mathbf{H}^{-1}.
\end{equation}

\begin{thm}\label{thm:sens_expression}
For the predicted residual $\mathbf{r}_{k} = \mathbf{y}_{m,k} - h\left( \mathbf{\hat x}_{k|k-1}, \varTheta_k \right)$, the fault sensitivity matrix is 	
\begin{equation}\label{eq:fsenmatrix}
\mathbf{S}_f = 
\left[ \begin{matrix}
\mathbf{\Phi} & \mathbf{I}
\end{matrix} \right]
\left[ \begin{matrix}
\mathbf{V} - \mathbf{J}_1 \mathbf{X} \mathbf{J}_1^\text{T} & \mathbf{0} \\
\mathbf{0} & \mathbf{I}
\end{matrix} \right]
\left[ \begin{matrix}
\mathbf{V}^{-1} & \mathbf{0} \\
-\mathbf{\Phi} & \mathbf{I}
\end{matrix} \right],
\end{equation}
with the definition of $\mathbf{\Phi}$ given in (\ref{eq:phi_def}).
\end{thm}

The proof is given in Appendix \ref{app:sens_respred}. 
Different from the averaged healthy measurements $\mathbf{\bar y}_{m,k}$ used in the MHE problem (\ref{eq:MHE}), the original output measurements $\mathbf{y}_{m,k}$ in (\ref{eq:ct_ss}) are used in residual generation. For the sake of simple notations, the complete output vector $\mathbf{y}_{m,k}$ is used. If the residual signal of particular sensor(s), e.g., AOA or VCAS, is of interest, then the corresponding rows of $\mathbf{r}_{k}$ in the above theorem are selected. In this case, all analysis in Section \ref{sect:fsen_MHEGR} remains the same except that $\mathbf{\Phi}$ changes according to the selected output components.

\subsection{Fault Sensitivity of Constrained-MHE-based Residual}\label{sect:fsen_CMHE}
When the faults are too small to activate any inequality constraints, fault sensitivity of CMHE-RG is the same as that of UMHE-RG. Next, we will show that the improved fault sensitivity of CMHE-RG is attributed to the active inequality constraints caused by sufficiently large faults. In this case, we have additional equalities $\mathbf{J}_a \Delta \mathbf{\hat z}_k = \mathbf{0}$ corresponding to the active constraints. Then the linearized KKT conditions (\ref{eq:KKTlin}) turn into 
\begin{equation}\label{eq:KKTlin2}
\left[ \begin{matrix}
\mathbf{H} & \mathbf{J}_2^\mathrm{T} & \mathbf{J}_a^\mathrm{T} \\
\mathbf{J}_2 & \mathbf{0} & \mathbf{0} \\
\mathbf{J}_a & \mathbf{0} & \mathbf{0} 
\end{matrix} \right] 
\left[ \begin{matrix}
\Delta \mathbf{\hat z}_k \\
\Delta \lambda \\
\Delta \mu_a
\end{matrix} \right] =
\left[ \begin{matrix}
\mathbf{J}_1^\mathrm{T} \mathbf{V}^{-1} \epsilon_k \\
\mathbf{0} \\
\mathbf{0}
\end{matrix} \right],
\end{equation}
where $\mu_a$ represents the Lagrange multiplier of the active inequality constraints.
The solution $\Delta \mathbf{\hat z}_k$ to (\ref{eq:KKTlin2}) and the fault sensitivity matrix $\mathbf{S}_f^a$ of its corresponding predicted residual are in the same form as (\ref{eq:Zsolution}) and (\ref{eq:fsenmatrix}), respectively:
\begin{align}
\Delta \mathbf{\hat z}_k &= \mathbf{X}_a \mathbf{J}_1^\mathrm{T} \mathbf{V}^{-1} \epsilon_k, \label{eq:Zsolution2} \\
\mathbf{S}_f^a &= \left[ \begin{matrix}
\mathbf{\Phi} & \mathbf{I}
\end{matrix} \right]
\left[ \begin{matrix}
\mathbf{V} - \mathbf{J}_1 \mathbf{X}_a \mathbf{J}_1^\text{T} & \mathbf{0} \\
\mathbf{0} & \mathbf{I}
\end{matrix} \right]
\left[ \begin{matrix}
\mathbf{V}^{-1} & \mathbf{0} \\
-\mathbf{\Phi} & \mathbf{I}
\end{matrix} \right], \label{eq:Sfa}
\end{align}
with
\begin{align}
\mathbf{X}_a &= \mathbf{H}^{-1} - \mathbf{H}^{-1} \mathbf{J}_{2a}^\mathrm{T} ( \mathbf{J}_{2a} \mathbf{H}^{-1} \mathbf{J}_{2a}^\mathrm{T} )^{-1} \mathbf{J}_{2a} \mathbf{H}^{-1}, \label{eq:Xa}\\
\mathbf{J}_{2a} &= \left[ \begin{matrix}
\mathbf{J}_{2}^\mathrm{T} & \mathbf{J}_{a}^\mathrm{T}
\end{matrix} \right]^\mathrm{T}. \label{eq:J2a}
\end{align}

\begin{thm}\label{thm:sens}
Assume that LICQ and sufficient second order condition hold before and after sensor faults occur, and additional inequality constraints become active in the presence of faults. With the same weighting matrix $\mathbf{V}$ in the MHE problem (\ref{eq:MHEcompact}), we have
$\mathbf{X}_a \leq \mathbf{X}$, hence $\mathbf{V} - \mathbf{J}_1 \mathbf{X}_a \mathbf{J}_1^\text{T} \geq \mathbf{V} - \mathbf{J}_1 \mathbf{X} \mathbf{J}_1^\text{T}$ which implies $\mathbf{S}_f^a (\mathbf{S}_f^a)^{\text{T}} \geq \mathbf{S}_f \mathbf{S}_f^\text{T}$, i.e., improved fault sensitivity of CMHE-RG compared to UMHE-RG. Besides, more activated inequality constraints lead to higher fault sensitivity.
\end{thm}

The proof is given in Appendix \ref{app:X_CMHE}.

\subsection{Trade-off between fault sensitivity and disturbance robustness}\label{sect:tune}
As pointed out in Section \ref{sect:sens_UMHE}, 
when the perturbation $\epsilon_k$ in the MHE problem (\ref{eq:MHEcompact}) represents the fault signal, $\mathbf{S}_f$ in (\ref{eq:fsenmatrix}) determines fault sensitivity. However, winds and VCAS sensor faults affect the same entries of the perturbation vector $\epsilon_k$, as pointed out in Section \ref{sect:prob}. When the perturbation $\epsilon_k$ comes from normal wind variations, $\mathbf{S}_f$ describes robustness to wind disturbances. Combining the above two aspects, it can be seen that higher fault sensitivity implies poorer disturbance robustness. 

First, we consider UMHE-RG, i.e., (\ref{eq:MHE})-(\ref{eq:MHEdef}) without inequality constraints. With fixed parameters $p_\alpha$, $q_\alpha$, and $\mathbf{R}$ that have been tuned for estimation performance, $p_d$ and $q_d$ are to be tuned for a trade-off between fault sensitivity and disturbance robustness. By using larger $p_d$ and $q_d$, more wind disturbances and a larger portion of fault perturbation $\epsilon_k$ can be interpreted by the assumed wind dynamics (\ref{eq:wind_dyn}), thus disturbance robustness improves 
but fault sensitivity decreases. Similarly, smaller $p_d$ and $q_d$ lead to loss of disturbance robustness and improvement of fault sensitivity.

According to Section \ref{sect:fsen_CMHE}, the active inequality contraints caused by faults in CMHE-RG bring benefits to the above performance trade-offs in the following two scenarios: 
\begin{enumerate}
	\item[a)] With fixed tuning parameters, CMHE-RG enables fault sensitivity improvement, and simultaneously maintains the same disturbance robustness if no inequality is activated by the normal wind variations.
	\item[b)] When increasing $p_d$ and $q_d$,  the disturbance robustness of both UMHE-RG and CMHE-RG improve. At the same time, however, UMHE-RG suffers from reduced fault sensitivity,  whereas fault sensitivity of CMHE-RG decreases only for small faults but increases for relatively larger faults. The reason is that the inequality constraints are still inactive in the presence of small faults, whereas some inequality constraints are activated by large faults if larger variations of the state estimates are allowed by larger $p_d$ and $q_d$.
\end{enumerate}

The assumed bounds on winds are critical for performance tradeoffs. 
If the assumed wind bounds are larger than the true bounds, both UMHE-RG and CMHE-RG have the same robustness to winds because no inequality constraints become active in the presence of real winds.
If the assumed wind bounds are smaller than the true bounds, the CMHE-RG would suffer from reduced robustness to winds because some inequality constraints are activated by large winds.
With any assumed wind bounds, the CMHE-RG achieves the same fault sensitivity as the UMHE-RG if the fault effect is too small to activate any inequality constraint, and it achieves higher fault sensitivity than the UMHE-RG as long as the fault effect results in some activated constraints.

\section{SIMULATION RESULTS AND COMPARISON}\label{sect:sim}
A high-fidelity Airbus civil aircraft simulator, developed in the RECONFIGURE project for numerical evaluation, is used \cite{Goup2014}. 
The considered scenario is level flight at the altitude of $\SI{5d3}{ft}$ with varying VCAS. Both horizontal and vertical winds $W_x $ and $W_z $ are simulated. Wind turbulence is also included in the simulator. 

The method proposed in Section \ref{sect:FTE} is applied to the longitudinal model described in Section \ref{sect:prob}.
The inequality constraints of the MHE problem (\ref{eq:MHE}) use the following bounds on wind speeds and accelerations: $\left| W_x \right| \leq \SI{120}{kts}$, $\left| W_z \right| \leq \SI{30}{kts}$ and $| \dot W_x | \leq \SI{15}{kts/s}$, $| \dot W_z | \leq \SI{15}{kts/s}$ 
Considering the scales of different variables, we select the weighting parameters in (\ref{eq:MHEdef}) as $q_\alpha = 10^{-8}$, $R_\alpha = 10^{-8}$, $R_{vz} = 2.5 \times 10^{-3}$, 
$R_{vc} = 2.5 \times 10^{-3}$.
In the arrival cost term, $p_\alpha = 10^{-6}$ is selected to be larger than $q_\alpha$, which means more belief on the AOA measurements than on the a priori AOA estimates.
For the augmented wind dynamics, the weighting parameters $p_d$ and $q_d$ are tuned for performance trade-offs as explained in Section \ref{sect:tune}.

The real-time iteration scheme described in \cite{Wan-IFAC2016} is implemented in the MATLAB2011b environment on a computer with a 3.4 GHz processor and 8 GB RAM. The horizon length of the MHE problem (\ref{eq:MHE}) is 5.  
The peak computational time per sample for CMHE-RG is 26 ms, below the sampling period of 40 ms.
The computational cost will be further reduced by structure-exploiting matrix manipulations and algorithmic approximations, e.g., lookup tables.

\begin{figure}[!h]
	\begin{center}
		\includegraphics[width=6cm]{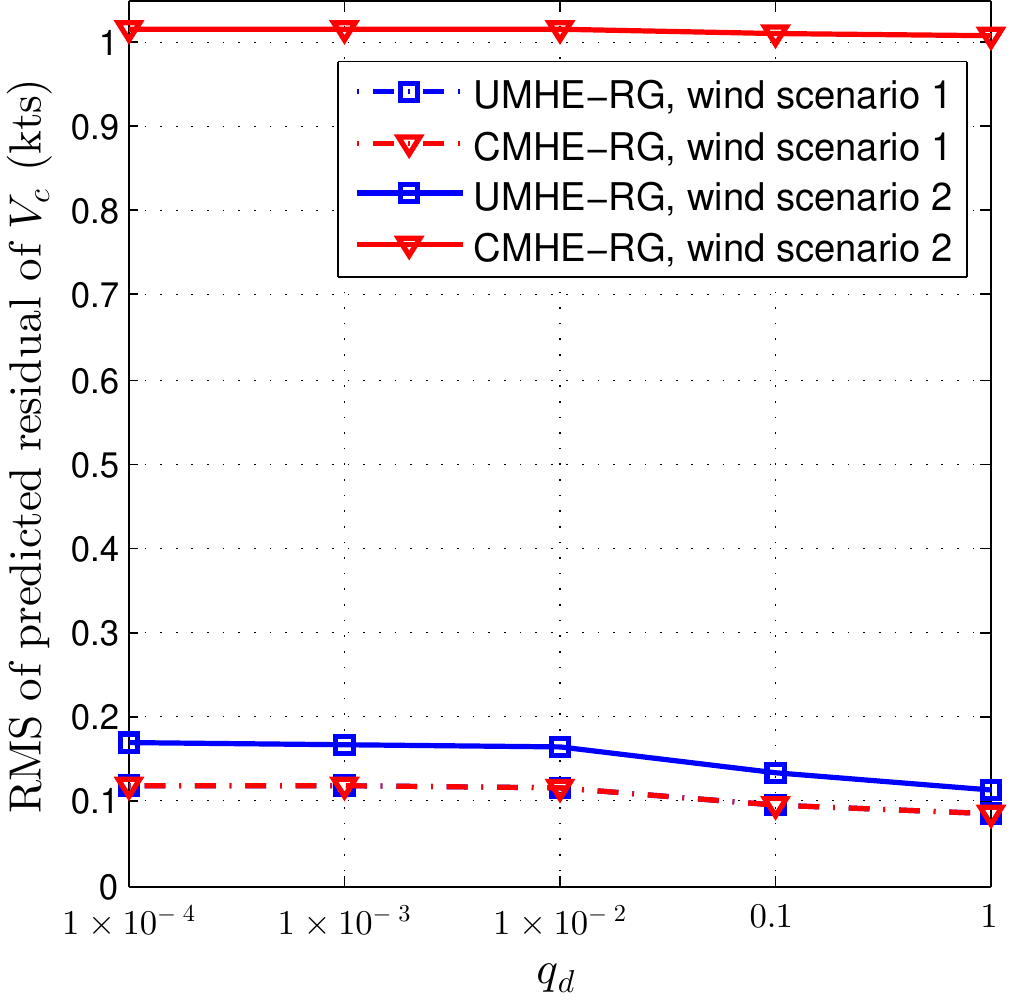}    
		\caption{Comparison of robustness to disturbances: RMS of the predicted residual with $p_d = 1$ and different $q_d$ in two fault-free scenarios. 1) $W_x$ varies from 0 to $\SI{10}{kts}$ with the acceleration $\SI{5}{kts/s}$, and $W_z$ varies from 0 to $\SI{-5}{kts}$ with the acceleration $\SI{-5}{kts/s}$; 2) $W_x$ varies from 0 to $\SI{21}{kts}$ with the acceleration $\SI{5}{kts/s}$, and $W_z$ is the same as in the first scenario.}
		\label{fig:est_perf}
	\end{center}
\end{figure}

In the following, we compare UMHE-RG and CMHE-RG in terms of disturbance robustness and fault sensitivity with different tuning parameters $p_d$ and $q_d$. For given wind disturbances, disturbance robustness can be measured by RMS of the predicted residual in the absence of faults. In this case, smaller RMS of the predicted residual implies higher robustness to disturbances. Therefore, to investigate disturbance robustness, we test the above two algorithms with $p_d = 1$ and different $q_d$ in the fault-free scenario. The performance comparisons with different $p_d$ are similar, thus are omitted. As shown by the wind scenario 1 in Figure \ref{fig:est_perf}, when both the wind speed and its acceleration are within their bounds used by CMHE-RG, the disturbance robustness of CMHE-RG is the same as that of UMHE-RG, because no inequality constraints are active when solving (\ref{eq:MHE}). 
In the wind scenario 2, however, the true wind $W_x$ is larger than its bound allowed by CMHE-RG. 
Then as analyzed in Sections \ref{sect:fsen_CMHE} and \ref{sect:tune}, the active constraints lead to more sensitivity, or equivalently, less robustness, to disturbances.  
This explains why CMHE-RG gives larger RMS of the predicted residual than UMHE-RG in wind scenario 2.

\begin{figure}[h]
	\begin{center}
		\includegraphics[width=7cm]{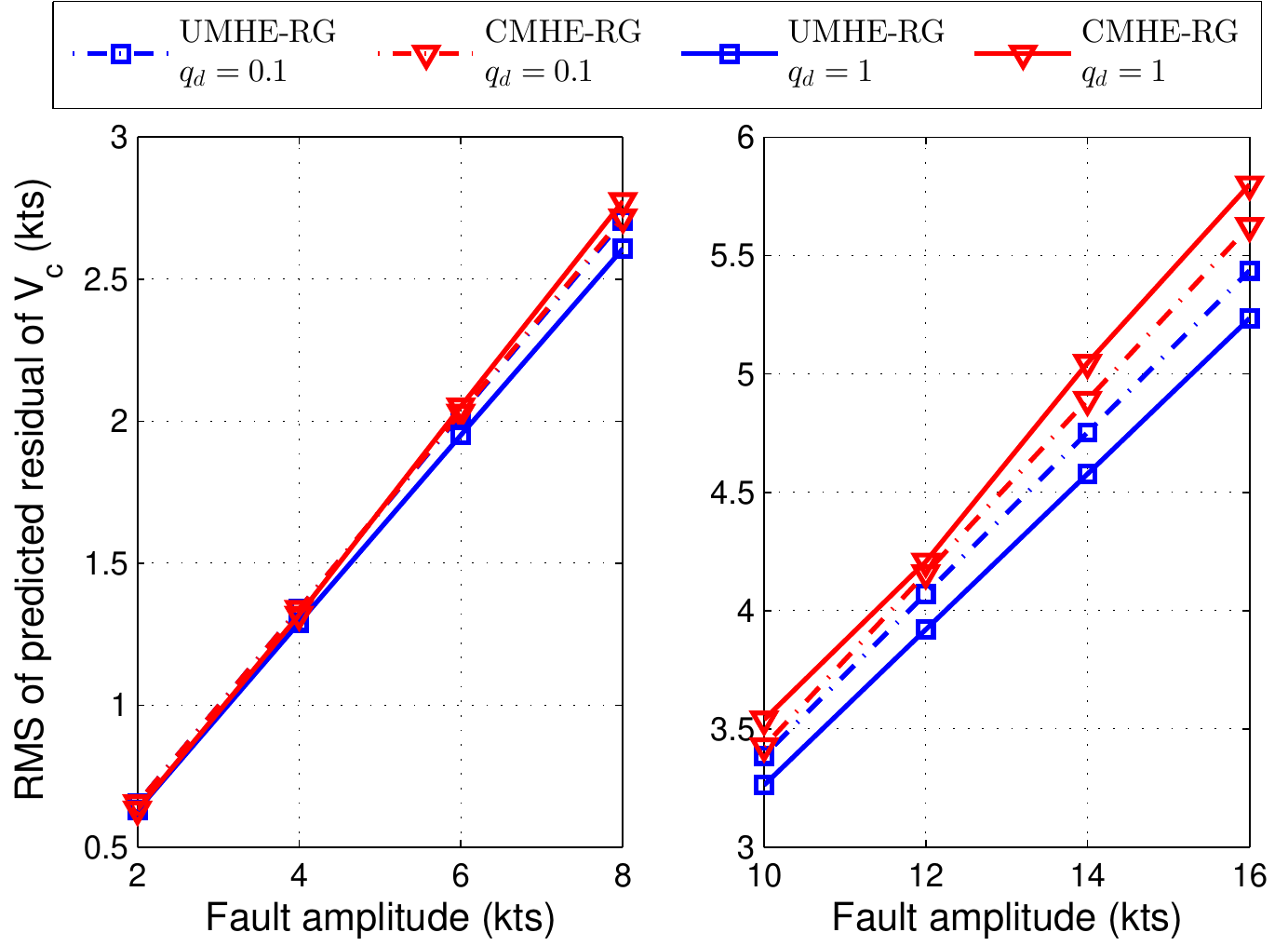}    
		\caption{Comparison of sensitivity to faults: RMS of predicted residual signal for different fault amplitudes in the presence of no wind}
		\label{fig:sens_perf}
	\end{center}
\end{figure}

Similarly, fault sensitivity of CMHE-RG is not directly evaluated by computing the fault sensitivity matrix $\mathbf{S}_f^a$ in (\ref{eq:Sfa}), because the active constraints required to compute $\mathbf{S}_f^a$ are unknown before solving the problem (\ref{eq:MHE}) at each time instant. 
Here, we indirectly evaluate fault sensitivity by RMS of the predicted residual within 100 samples immediately after fault injection. This indirect evaluation requires excluding the wind effect in the predicted residual, thus no wind is included in the simulations for evaluating fault sensitivity. In this case, larger RMS of the predicted residual implies higher sensitivity to faults. As explained in the last paragraph of Section \ref{sect:prob}, fault sensitivity is more critical to VCAS sensor fault than to AOA sensor faults. Hence we focus on the constant bias fault in one VCAS sensor for comparisons. 
The simulation results are shown in Figure \ref{fig:sens_perf}. For fault amplitude smaller than $\SI{5}{kts}$, CMHE-RG produces the same RMS of predicted residual as UMHE, i.e., both algorithms have the same fault sensitivity, and have their fault sensitivity slightly decreased when $q_d$ increases from 0.1 to 1. 
For fault amplitude larger than $\SI{5}{kts}$, CMHE-RG gives larger RMS of predicted residual, which implies higher fault sensitivity, than UMHE-RG given either $q_d=0.1$ or $q_d=1$. Moreover, for these larger faults,
when $q_d$ increases from 0.1 to 1, fault sensitivity of UMHE reduces, whereas fault sensitivity of CMHE  increases.
These above observations can be explained by no active inequality constraints in the presence of faults smaller than $\SI{5}{kts}$, and additional active constraints due to larger faults, which are consistent with Theorem \ref{thm:sens} and the two scenarios given at the end of Section \ref{sect:tune}.

\begin{figure}[h]
	\begin{center}
		\includegraphics[width=6cm]{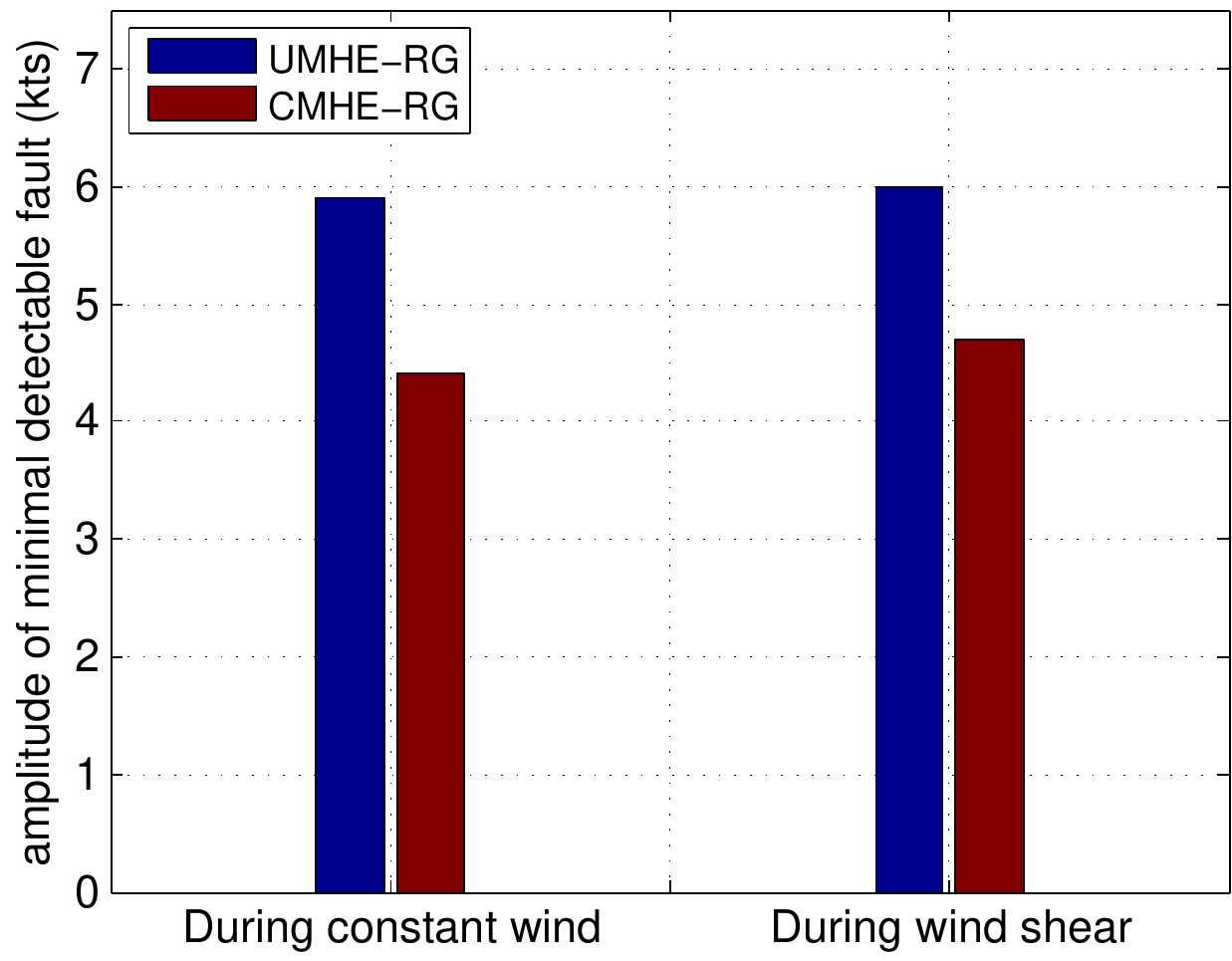}    
		\caption{Comparison of minimal detectable VCAS sensor bias during constant winds ($W_x=W_z=\SI{10} {kts}$ ) and wind shear ($\dot W_x= \dot W_z = \SI{10}{kts/s}$)}
		\label{fig:minf}
	\end{center}
\end{figure}


\begin{figure}[!h]
	\centering
	\subfigure[Measurements and winds]{
		\includegraphics[width=8cm]{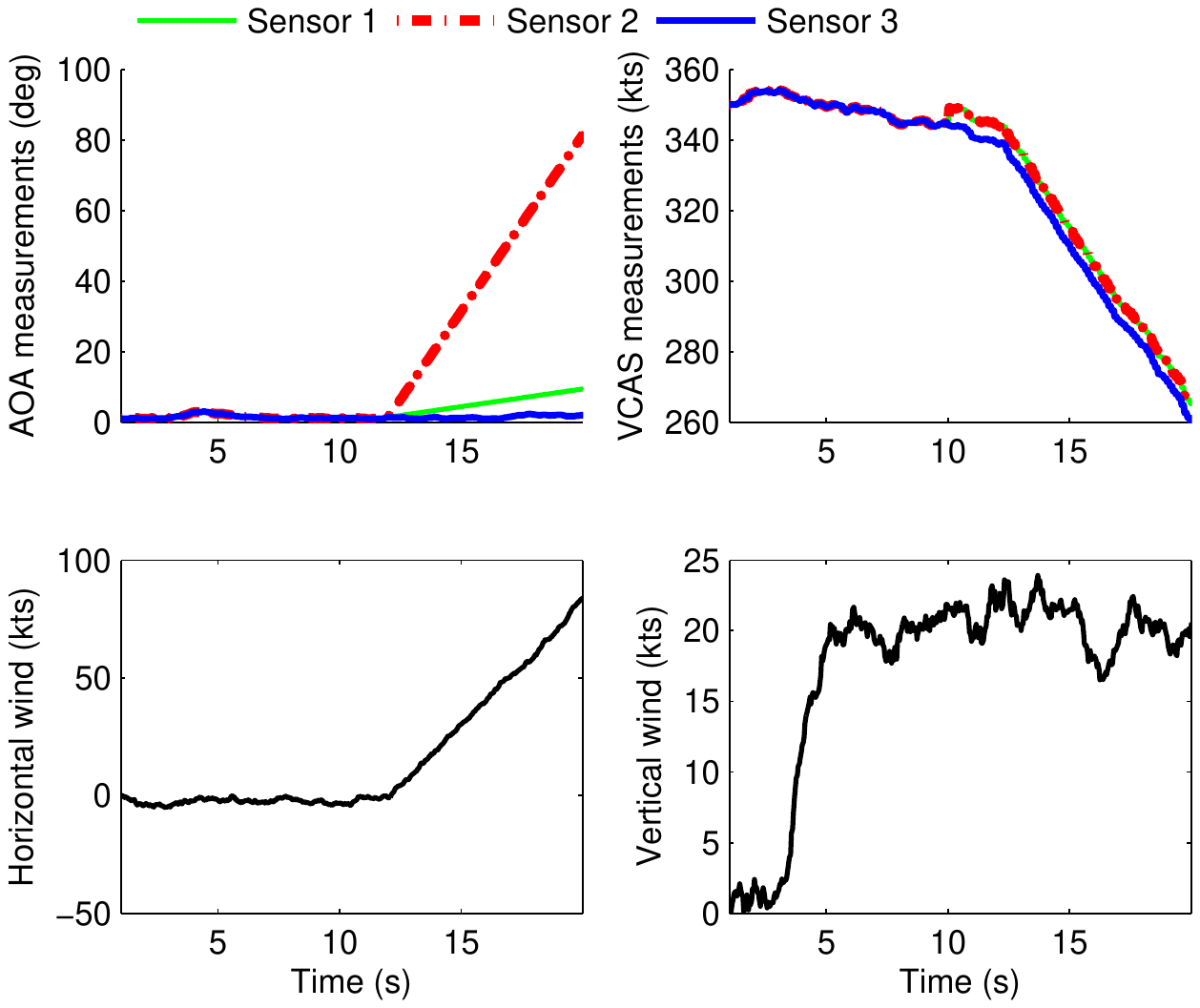}\label{fig:winds}	
	}
	\subfigure[AOA and VCAS estimates]{
		\includegraphics[width=8cm]{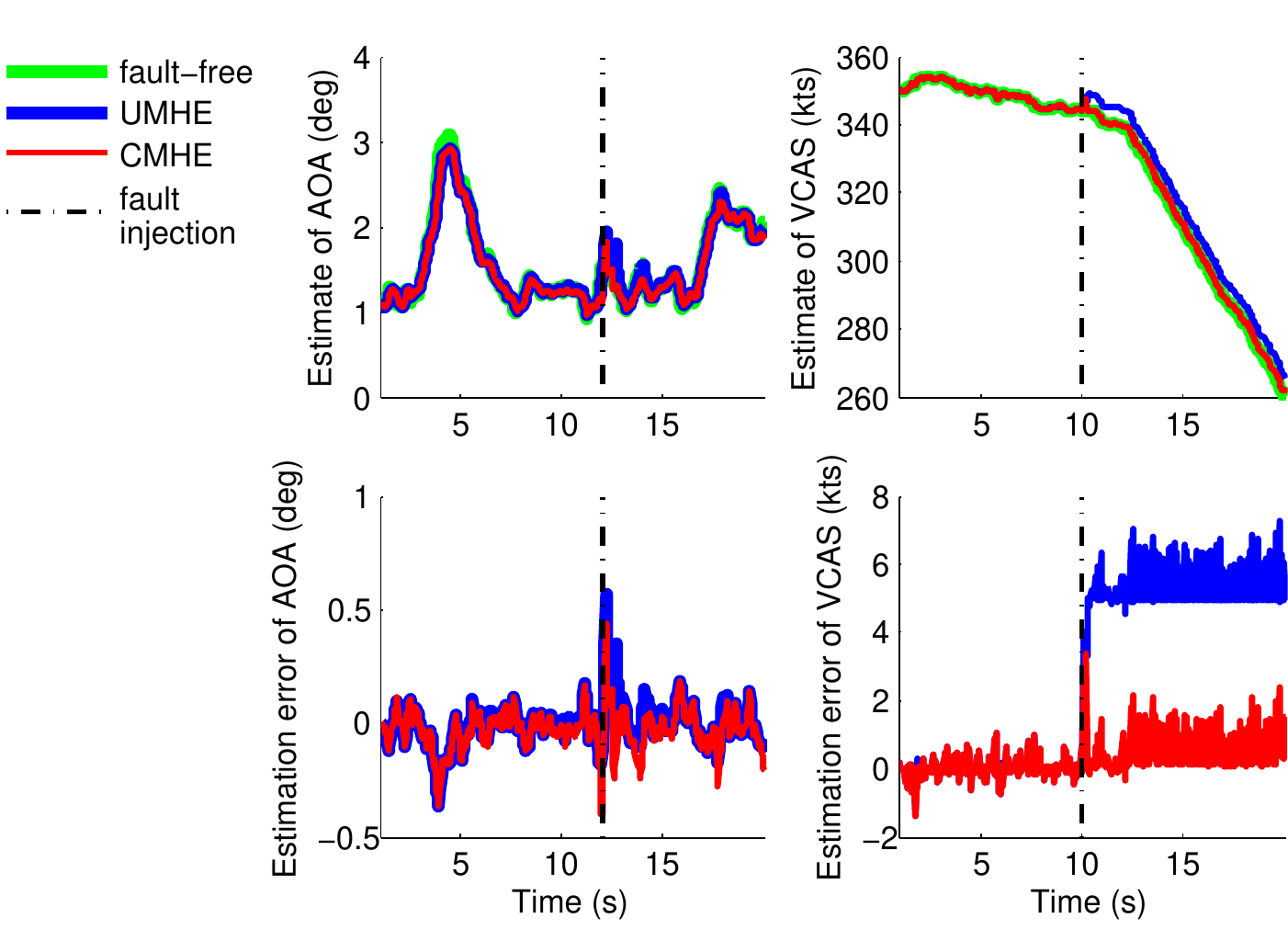}\label{fig:FDI_est}
	}
	\subfigure[FDI of AOA and VCAS sensors]{
		\includegraphics[width=8cm]{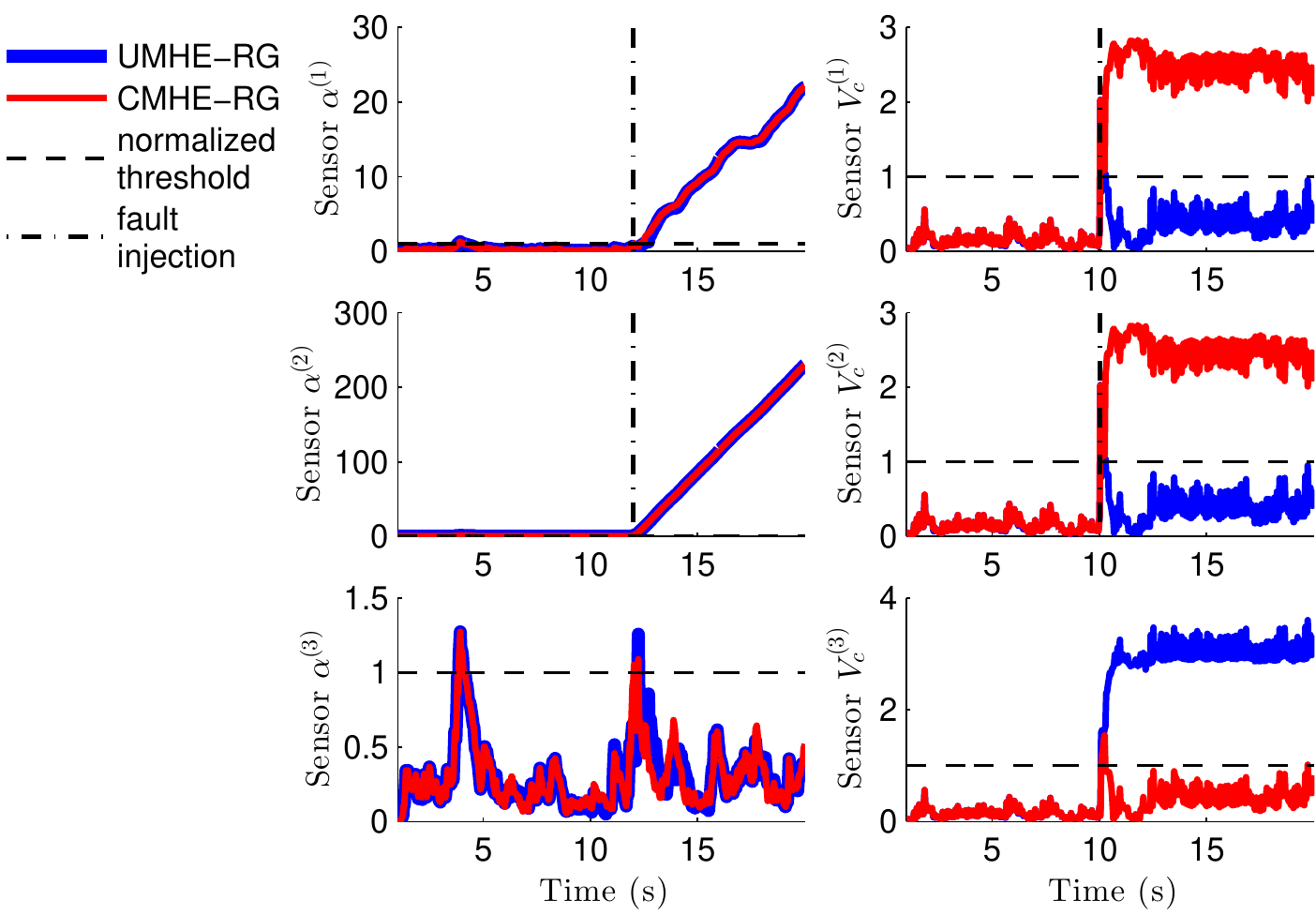}\label{fig:FDI_res}	
	}
	\caption{Comparison of estimation and FDI results for simultaneous AOA and VCAS sensor faults in the presence of wind and turbulence (tuning parameters $p_d = 1$ and $q_d=1$)}
	\label{fig:FTEresult}
\end{figure}

By increasing the bias amplitude of VCAS sensor by $\SI{0.1}{kts}$ each time in a sequence of simulations, we find the approximate size of the minimal detectable faults of the UMHE-RG and CMHE-RG. 
Smaller minimal detectable fault implies higher fault sensitivity.
In the FDI logic, we select the length of the residual evaluation window to be $N_{\text{eval}}=10$. 
For fair comparison, the detection thresholds $J_{\text{th}}$ in (\ref{eq:FDlogic}) of the two algorithms need to be carefully chosen. The fault-free simulation data in the presence of wind with largest speed ($W_x=\SI{20}{kts}$) and fastest acceleration ($\dot W_x=\SI{15}{kts/s}$) within the assumed bounds are used to determine the thresholds, 
and each threshold is set to be the smallest value that ensures zero false alarms for each algorithm. 
A VCAS sensor is concluded to be faulty when its residual evaluation is above the detection threshold for at least three times in the past 10 time instants.
Figure \ref{fig:minf} shows the results during constant winds and wind shear. The CMHE-RG reduces the amplitude of minimal detectable fault in both scenarios.

Finally, we consider simultaneous bias of VCAS sensors and runaway of AOA sensors: constant bias $\SI{5}{kts}$ in sensor $V_{c}^{(1)}$, constant bias $\SI{7}{kts}$ in sensor $V_{c}^{(2)}$, runaway fault at rate $\SI{1}{deg/s}$ in sensor ${\alpha}^{(1)}$, runaway fault at rate $\SI{10}{deg/s}$ in sensor ${\alpha}^{(2)}$, as plotted in Figure \ref{fig:winds}. 
The VCAS/AOA estimates and the residual evaluations for all sensors are illustrated in Figures \ref{fig:FDI_est} and \ref{fig:FDI_res}. It can be seen that both methods can correctly isolate AOA sensor faults, but CMHE-RG gives better performance than UMHE-RG in isolating VCAS sensor faults: UMHE-RG gives false alarms on the healthy sensor $V_{c}^{(3)}$, while CMHE-RG isolates all VCAS sensor faults within less than $\SI{0.2}{s}$. 

\section{CONCLUSIONS AND FUTURE WORKS}\label{sect:con}
This paper presented a moving horizon estimation based approach for robust air data sensor fault diagnosis. The challenge due to simultaneous influence of winds and sensor faults on calibrated airspeed measurements is tackled by incorporating wind dynamics and exploiting wind bounds in residual generation. The Karush-Kuhn-Tucker condition of the formulated moving horizon estimation problem is analyzed to show 
that the constrained residual generator based on moving horizon estimation has improved fault sensitivity because some inequality constraints become active in the presence of faults. 
With tuning parameters that increases disturbance robustness, conventional unconstrained residual generators would simply lose fault sensitivity, whereas our proposed constrained residual generator gains additional fault sensitivity as long as the fault activates inequality constraints. A high-fidelity Airbus simulator is used to illustrate the advantage of our proposed approach. 

It should be pointed out that sensitivity to fault and robustness to disturbance are indirect measures related to fault detection rate and false alarm rate. Besides theoretically statistical analysis, extensive Monte Carlo evaluations of our proposed approach and comparisons with difference approaches are the focus of our future work.

%



\bibliographystyle{plain}
\bibliography{bibACC2016}

\appendices
\section{Modeling of Longitudinal Motions}\label{app:model}
Considering wind speeds $W_x$ and $W_z$ in the inertial frame, the longitudinal dynamics of AOA in the body frame is
\begin{equation}\label{eq:al_dyn_exact}
\begin{aligned}
\dot{\alpha} = \frac{1}{V_t} f_{\alpha} (\alpha, \varTheta) + q + \frac{1}{V_t} f_w (\alpha, \mathbf{w}, \varTheta),
\end{aligned}
\end{equation}
with 
\begin{align}
& f_{\alpha} (\alpha, \varTheta) = n_z \cos \alpha - n_x \sin \alpha + g \cos (\alpha - \theta), \\
& f_w (\alpha, \mathbf{w}, \varTheta) = \dot{W}_x sin(\alpha - \theta) - \dot{W}_z cos(\alpha - \theta).
\end{align}
The above model is obtained from the third row of the equation (1) of \cite{Lee2013} by (a) including only longitudinal motions, and (b) equivalently replacing the terms of aerodynamic and propulsion forces $\mathbf{R}_{\mathbf{B}_\mathbf{T}\mathbf{W}_\mathbf{T}} A_T + P_T$ in \cite{Lee2013} with load factors $[\begin{matrix}
n_x & 0 & n_z
\end{matrix}]^\text{T}$, as in the equation (14) of \cite{Eykeren2014}.
Note that measurements of true airspeed $V_t$ are unreliable in the presence of VCAS sensor fault \cite{Goup2014,Goup2015}, thus cannot be directly used in the model (\ref{eq:al_dyn_exact}).
In this case, we make the following approximations by using reliable measurements of ground speed $V_g$ to replace $V_t$ in (\ref{eq:al_dyn_exact}). 
Let $\Delta V = V_t - V_g$ denote the difference between true airspeed $V_t$ and ground speed $V_g$ due to the winds. Since we have 
\begin{equation}\label{eq:approx}
V_g^2 \gg \Delta V  f_{\alpha} (\alpha, \varTheta), 
V_t \gg \dot W_x, V_t \gg \dot W_z
\end{equation}
for the flight scenarios in the RECONFIGURE project \cite{Goup2014,Goup2015}, the approximations 
\begin{subequations}
\begin{align}\label{eq:xx}
& \frac{1}{V_t} f_{\alpha} (\alpha, \varTheta) \approx \left( \frac{1}{V_g} - \frac{1}{V_g^2} \Delta V \right) f_{\alpha} (\alpha, \varTheta) \approx \frac{1}{V_g} f_{\alpha} (\alpha, \varTheta), \\
& \frac{1}{V_t} f_w (\alpha, \mathbf{w}, \varTheta) \approx 0
\end{align}
\end{subequations}
can be used to derive the following simplified model from (\ref{eq:al_dyn_exact}) without sacrificing FDI and estimation performance:
\begin{equation}\label{eq:al_dyn_app}
\begin{aligned}
\dot{\alpha} &= \frac{1}{V_g} f_{\alpha} (\alpha, \varTheta) + q + u_{\alpha}.
\end{aligned}
\end{equation}   
Here the first two terms on the right-hand side are represented by $f (\alpha, \varTheta)$ in the model (\ref{eq:ct_ss}), and $u_{\alpha}$ accounts for the model mismatch including the approximation errors in (\ref{eq:xx}) and the effect of noises in the measured parameter $\varTheta$. 

The output equation for vertical speed $V_z$ in the inertial frame is 
(Equation (2.4-5) of \cite{Stevens1992})
\begin{equation}\label{eq:Vz}
\begin{aligned}
V_z &= -V_t \sin(\alpha - \theta) + W_z \\ 
&= - h_{vt} (\alpha, \mathbf{w}, \varTheta) \sin(\alpha - \theta) + W_z.
\end{aligned}
\end{equation}
To avoid using the unreliable measurements of $V_t$ mentioned in the previous paragraph, the function 
\begin{equation*}
\begin{aligned}
h_{vt} (\alpha, \mathbf{w}, \varTheta) &= - W_x \cos(\alpha-\theta) + W_z \sin(\alpha-\theta) \\
+& \sqrt{ V_g^2 - \left[ W_x \sin(\alpha-\theta) + W_z \cos(\alpha-\theta) \right]^2 } 
\end{aligned}
\end{equation*}
in (\ref{eq:Vz}) is used to transform the ground speed $V_g$ into the true airspeed $V_t$, which is constructed based on (Equation (1.5-6) of \cite{Stevens1992})
\begin{equation*}
\left\{ \begin{array}{l}
V_g^2 = {u_g^2 + w_g^2} \\
u_g = V_t \cos{\alpha} + W_x \cos{\theta} + W_z \sin{\theta} \\
w_g = V_t \sin{\alpha} + W_x \sin{\theta} - W_z \cos{\theta}.
\end{array} \right.
\end{equation*}

The output equation for fault-free VCAS $V_c$ consists of two conversions: 1) from ground speed $V_g$ to true airspeed $V_t$ by the function $h_{vt} (\alpha, \mathbf{w}, \varTheta)$; and 2) from $V_t$ to $V_c$ \cite{Davies2003,Hardier2013}, i.e.,
\begin{equation}\label{eq:vc_vt}
\begin{aligned}
V_c &= \sqrt {5\gamma R{T_0}} \varrho (V_t, T, \bar p),   \\
&= \sqrt {5\gamma R{T_0}} \varrho \left( h_{vt} (\alpha, \mathbf{w}, \varTheta), T, \bar p \right)
\end{aligned}
\end{equation}
with $T = {T_0} + Lz$, $\bar p = {\left( {1 + \frac{L}{{{T_0}}}z} \right)^{\frac{g}{{ - RL}}}}$,
\begin{align*}
& \varrho(V_t, T, \bar p) = \sqrt{ {{{\left[ {\left( {{{\left( {1 + \frac{{V_t^2}}{{5\gamma RT}}} \right)}^{3.5}} - 1} \right)\bar p + 1} \right]}^{\frac{1}{3.5}}} - 1} },
\end{align*}
where $z$, $T$, and $\bar p$ represent pressure altitude, outside air temperature, static pressure scaled by the ground static pressure value, respectively. The constants $T_0$, $L$, $R$, and $\gamma$ take their values according to International Standard Atmosphere \cite{Davies2003}: $T_0 = 288.15 K$, $L = -6.5 K/\text{km}$, $R = 287.05287 \, (\text{m/s})^2 \cdot K$, and $\gamma = 1.4$.
$z$ in $\varTheta$, $T$ and $\bar p$ in (\ref{eq:vc_vt}) uses altitude measurements.

\section{Proof of Theorem \ref{thm:sens_expression}}\label{app:sens_respred}
%

Let $\mathbf{\hat x}_{k|k} = \mathbf{P}_s \mathbf{\hat z}_{k} (\mathcal{I}_k)$ (with $\mathbf{P}_s = \left[ \begin{matrix}
\mathbf{0} & \cdots & \mathbf{0} & \mathbf{I}
\end{matrix} \right]$), and $\mathcal{\hat I}_k = F_1 \left( \mathbf{\hat z}_k (\mathcal{I}_k) \right)$, then we have 
$\mathbf{\hat z}_k (\mathcal{I}_k) = \mathbf{\hat z}_k (\mathcal{\hat I}_k)$ according to (\ref{eq:MHEcompact}). From (\ref{eq:dt_longoutput}), (\ref{eq:Zsolution}), and (\ref{eq:MHEcon}), the one-step-ahead output prediction $\mathbf{\hat y}_{k|k-1}$ can be written as
\begin{equation}\label{eq:out_est}
\begin{aligned}
\mathbf{\hat y}_{k|k-1} &= h\left( \mathbf{\hat x}_{k|k-1}, \varTheta_k \right) \\
&= h \left( F( \mathbf{\hat x}_{k-1|k-1}, \mathbf{0}, \varTheta_{k-1}), \varTheta_k \right) \\
&= h \left( F( \mathbf{P}_s \mathbf{\hat z}_{k-1} (\mathcal{\hat I}_{k-1}), \mathbf{0}, \varTheta_{k-1}), \varTheta_k \right) \\
&= \nu \left( \mathcal{\hat I}_{k-1}, \varTheta_{k-1}, \varTheta_k \right) \\
&= \nu \left( F_1 \left( \mathbf{\hat z}_{k-1} (\mathcal{I}_{k-1}) \right), \varTheta_{k-1}, \varTheta_k \right).
\end{aligned}
\end{equation}
In the above equation, the function $\nu (\cdot)$ describes how the output prediction relies on the estimation of the past information vector, and we define 
\begin{equation}\label{eq:phi_def}
\mathbf{\Phi} = \tfrac{\partial \nu }{\partial \mathcal{\hat I}_{k-1}}.
\end{equation}
With (\ref{eq:info_vector}), (\ref{eq:Zsolution}) and (\ref{eq:out_est}), the sensitivity of the predicted residual $\mathbf{r}_{k} = \mathbf{y}_{m,k} - h\left( \mathbf{\hat x}_{k|k-1}, \varTheta_k \right)$ to the fault perturbation $\epsilon_k$ can be analyzed via first-order Taylor expansion as below:
\begin{equation}\label{eq:res_analysis}
\begin{aligned}
\mathbf{r}_{k} 
&= \mathbf{y}_{m,k} - \nu \left( F_1 \left( \mathbf{\hat z}_{k-1} (\mathcal{I}_{k-1}^0 + \epsilon_{k-1}) \right), \varTheta_{k-1}, \varTheta_k \right) \\
& \approx \mathbf{y}_{m,k} - \nu \left( F_1 \left( \mathbf{\hat z}_{k-1} (\mathcal{I}_{k-1}^0 ) \right) + J_1 \Delta \mathbf{\hat z}_{k-1}, \varTheta_{k-1}, \varTheta_k \right) \\
&\approx \mathbf{y}_{k} + \mathbf{n}_k + \mathbf{f}_{k} - \nu \left( F_1 \left( \mathbf{\hat z}_{k-1} (\mathcal{I}_{k-1}^0 ) \right), \varTheta_{k-1}, \varTheta_k \right) \\
& \quad\, - \mathbf{\Phi} \mathbf{J}_1 \mathbf{X} \mathbf{J}_1^\mathrm{T} \mathbf{V}^{-1} \epsilon_{k-1} \\
&= {\mathbf{y}_{k} + \mathbf{n}_k - \nu \left( F_1 \left( \mathbf{\hat z}_{k-1} (\mathcal{I}_{k-1}^0 ) \right), \varTheta_{k-1}, \varTheta_k \right)} \\
& \quad\, + \left[ \begin{matrix}
- \mathbf{\Phi} \mathbf{J}_1 \mathbf{X} \mathbf{J}_1^\text{T} \mathbf{V}^{-1} & 
\mathbf{I}
\end{matrix} \right]
\left[ \begin{matrix}
\epsilon_{k-1} \\
\mathbf{f}_k
\end{matrix} \right]
\end{aligned}
\end{equation}
Note that $\mathbf{\Phi}$ is defined in (\ref{eq:phi_def}), and its value in the above equation is not related to faults, but determined by $F_1 \left( \mathbf{\hat z}_{k-1} (\mathcal{I}_{k-1}^0 ) \right)$, i.e., the fault-free information vector $\mathcal{I}_{k-1}^0$.
Hence the last term of the last equation in (\ref{eq:res_analysis}) shows the effect of faults on the predicted residual, and the fault sensitivity matrix is 
$$
\mathbf{S}_f = \left[ \begin{matrix}
- \mathbf{\Phi} \mathbf{J}_1 \mathbf{X} \mathbf{J}_1^\text{T} \mathbf{V}^{-1} & 
\mathbf{I}
\end{matrix} \right]
$$
which can be rewritten as (\ref{eq:fsenmatrix}).

\section{Proof of Theorem \ref{thm:sens}}\label{app:X_CMHE}
Let the symmetric matrix $\mathbf{\Pi}$ denote the matrix square root of the Hessian matrix $\mathbf{H}$, i.e., $\mathbf{H} = \mathbf{\Pi} \cdot \mathbf{\Pi}$, and define
\begin{align}
\mathcal{P} &= \mathbf{I} - \mathbf{\Pi}^{-1} \mathbf{J}_2^\mathrm{T} \left( \mathbf{J}_2 \mathbf{\Pi}^{-2} \mathbf{J}_2^\mathrm{T} \right)^{-1} \mathbf{J}_2 \mathbf{\Pi}^{-1}, \label{eq:Pcal}\\
\mathcal{P}_a &= \mathbf{I} - \mathbf{\Pi}^{-1} \mathbf{J}_{2a}^\mathrm{T} \left( \mathbf{J}_{2a} \mathbf{\Pi}^{-2}  \mathbf{J}_{2a}^\mathrm{T} \right)^{-1} \mathbf{J}_{2a} \mathbf{\Pi}^{-1}. \label{eq:Pacal}
\end{align}
Then $\mathbf{X}$ in (\ref{eq:X}) and $\mathbf{X}_a$ in (\ref{eq:Xa}) can be rewritten as
\begin{align}\label{eq:XXa}
\mathbf{X}= \mathbf{\Pi}^{-1} \mathcal{P} \mathbf{\Pi}^{-1} \; \text{and} \; 
\mathbf{X}_a= \mathbf{\Pi}^{-1} \mathcal{P}_a \mathbf{\Pi}^{-1},
\end{align}
respectively. 
Let $\mathcal{N}(\cdot)$ denote the left null space of a matrix.
It can be seen from (\ref{eq:Pcal}) and (\ref{eq:Pacal})
that $\mathcal{P}$ and $\mathcal{P}_a$ are two orthogonal projectors onto the left null spaces 
$\mathcal{N}(\mathbf{J}_2 \mathbf{\Pi}^{-1})$ and $\mathcal{N}(\mathbf{J}_{2a} \mathbf{\Pi}^{-1})$, respectively.
According to (\ref{eq:J2a}), the left null space $\mathcal{N}(\mathbf{J}_{2a} \mathbf{\Pi}^{-1})$ is a subset of $\mathcal{N}(\mathbf{J}_{2} \mathbf{\Pi}^{-1})$, which implies
$\mathcal{P}_a < \mathcal{P}$. Therefore, $\mathbf{X}_a \leq \mathbf{X}$ and $\mathbf{V} - \mathbf{J}_1 \mathbf{X}_a \mathbf{J}_1^\text{T} \geq \mathbf{V} - \mathbf{J}_1 \mathbf{X} \mathbf{J}_1^\text{T}$ according to (\ref{eq:XXa}). Then it can be concluded from (\ref{eq:fsenmatrix}) and (\ref{eq:Sfa}) that $\mathbf{S}_f^a (\mathbf{S}_f^a)^\mathrm{T} \geq \mathbf{S}_f \mathbf{S}_f^\mathrm{T}$.

For the same reason as above, the left null space $\mathcal{N}(\mathbf{J}_{2a} \mathbf{\Pi}^{-1})$ with more active inequality constraints in $\mathbf{J}_{2a}$ is a subset of $\mathcal{N}(\mathbf{J}_{2a} \mathbf{\Pi}^{-1})$ with fewer active inequality constraints in $\mathbf{J}_{2a}$. 
Hence when more inequality constraints are active in solving the MHE problem, $\mathcal{P}_a$ becomes smaller, and fault sensitivity increases accordingly.

\end{document}